\documentclass{icrc}

\usepackage{times}
\usepackage{graphicx} 

\def\gr{$\gamma$-ray }
\def\grs{$\gamma$-rays }
\def\grsn{$\gamma$-rays}
\def\be{\begin{equation}}
\def\ee{\end{equation}}
\def\part#1#2{{{\partial #1}\over {\partial #2}}}
\begin{document}

\title{Gamma Ray Astronomy}
\author{Martin Pohl}
\affil{Institut f\"ur Theoretische Physik, Lehrstuhl 4, 
Ruhr-Universit\"at Bochum, 
44780 Bochum, Germany}
\correspondence{M. Pohl (mkp@tp4.ruhr-uni-bochum.de)}

\firstpage{1}
\pubyear{2001}


\maketitle

\begin{abstract}
This paper summarizes recents results in \gr astronomy, most of which were
derived with data from ground-based \gr detectors. Many of the contributions 
presented at this conference involve multiwavelength studies which combine
ground-based \gr measurements with optical data or space-based X-ray and 
\gr measurements. Besides measurements of the diffuse emission from
the Galaxy, observations of blazars, \gr bursts, and supernova remnants
this paper also covers theoretical models for the acceleration of radiating 
particles and their emission mechanisms in these sources.
\end{abstract}

\section{Introduction}

To date, \gr astronomy is a rapidly evolving field, the clearest indication
of which is the fact that a many reports in the AGN session were rewritten
to include spectacular results obtained only a few weeks prior to the start
of this conference.

This rapporteur paper covers 192 contributed papers on a range of topics in 
\gr astronomy drawn from the sessions OG 2.x. The classes of objects covered
include supernova remnants (OG 2.2), pulsars and plerions (OG 2.2), active galactic nuclei (OG 2.3), Galaxy Clusters (OG 2.3), \gr bursts (OG 2.4).
Also presented were results on diffuse \gr emission (OG 2.1) and on
astroparticle physics (OG 2.7). Briefly summarized are reports on projects, instrumentation, and analysis techniques (OG 2.5).

\section{Projects and instrumentation}

Since the de-orbiting of the Compton Gamma-Ray Observatory (CGRO) in June 2000
no space-based \gr detector has been operational. During the Nineties
the Energetic Gamma-Ray 
Experiment Telescope (EGRET) aboard CGRO was very successful in detecting
GeV \grs from around 70 AGN, 8 pulsars, and 170 sources not yet
identified firmly with known objects \citep{3eg}. EGRET
has also measured the spectrum
and the spatial distribution of the diffuse galactic \gr emission 
with unprecedented sensitivity and resolution \citep{hu97}.

The planned successor to EGRET, the Gamma-ray Large Area Space Telescope 
(GLAST), will not be launched before the year 2006. GLAST 
will offer a factor of eight more sensitive area to \grs than did EGRET, which 
combined with a much larger field-of-view and a better energy and spatial
resolution will provide a sensitivity gain by a factor of thirty 
compared with EGRET \citep{od-ic}.

During the time before GLAST becomes operational, two initiatives in
satellite-based \gr astronomy are planned to provide astronomical data
at GeV energies. The italian AGILE satellite (Astro-rivelatore Gamma a
Immagini LEggero) will offer a sensitive area
similar to that of EGRET and an angular resolution somewhat better than
EGRET \citep{mt01}. The sensitivity of AGILE as a \gr detector will however
be compromised by its limited energy resolution. The Alpha Magnetic
Spectrometer (AMS), an instrument originally devised for the search for
antimatter in cosmic rays, will also be able to identify \grs \citep{bl-ic}.
For technical reasons AMS will mostly detect \grs with energies of a few 
GeV or higher, albeit with a sensitive area and an angular resolution
slightly superior to those of EGRET. AMS can be expected to significantly
contribute to our understanding of diffuse Galactic \gr emission, most
notably the GeV excess, but may suffer from limited statistic in studies
of point sources.

All GeV \gr experiments use pair production in thin foils of high-$Z$ material
to actually detect the \grsn. Different techniques are used to track the 
$e^+/e^-$-pairs and to measure their energy, though. In principle the
\gr energy threshold is around 10 MeV, but the short range of the pairs and
small-angle scattering in the tracker significantly deteriorate
the detector performance below 100 MeV. Towards high \gr energies self-vetoing
and the finite thickness of the calorimeter can reduce the quality of 
measurement. The main problems with satellite-based \gr detectors, however,
are the technical constraints which prohibit satellite payloads with an
effective area of much more than a squaremeter. The flux of all cosmic
\gr sources falls off with photon energy and therefore 
the scientific return of the
\gr detectors at high photon energies is limited by statistics rather
than inapplicability of the technique of measurement. GLAST will have an 
effective high energy limit of a few hundred GeV.

Photons with energies of a hundred GeV or higher generate electromagnetic 
showers in the earth atmosphere. The secondary particles thus produced 
move faster than the phase velocity of electromagnetic light and therefore
emit optical \v Cerenkov-light that can be measured with suitable telescopes.
Existing imaging \v Cerenkov telescopes
such as WHIPPLE \citep{cawl90}, CAT \citep{barrau98},
HEGRA \citep{daum97}, CANGAROO \citep{hara93}, or TACTIC \citep{bhatt-ic}
have energy thresholds between 300 GeV and 2 TeV, but a 
sensitive area $\ge 10^3\ {\rm m^2}$, because the atmosphere is used 
as the interaction site. Forthcoming or planned experiments will have lower
energy thresholds and hence a higher sensitivity than the existing 
installations. Commencing operations in 2002, MAGIC \citep{lorenz-ic}
and VERITAS \citep{buck-ic} will observe the northern hemisphere
while H.E.S.S. \citep{hofm-ic} and CANGAROO III \citep{mori-ic}
will study the southern sky. The MAGIC project will use a single large
telescope optimized to provide a low energy threshold of 10--30 GeV,
whereas the other three observatories will use multiple telescopes to
simultaneously measure the \v Cerenkov-light of a shower, thus providing
a very good hadron rejection and an excellent energy resolution.

Non-imaging observatories such as CELESTE 
\citep{dena01}, STACEE \citep{cova-ic}, Solar-2 \citep{tume-ic}, GRAAL 
\citep{arqu-ic}, and PACT \citep{chit-ic} are now becoming operational. 
The very large mirror area of these observatories 
makes for a very low energy threshold of less than 30 GeV with large
effective area, but the shower reconstruction and the
hadron rejection are more difficult than for imaging observatories, for only
the arrival time of the \v Cerenkov shower front can be measured. 

All these \v Cerenkov telescopes provide a very good point source sensitivity,
which allows to measure source variability on time scales of around one 
hour \citep{khel-ic,feg-ic}. However, only one source can be observed at a time,
and therefore the actual measurement of interesting behaviour of a source
requires either luck or an indication of activity from other resources.
The atmospheric
\v Cerenkov telescopes are thus rather complementary to GLAST or to monitoring 
devices in the TeV energy range such as MILAGRO, which experiment detects
\grs from a field nearly $2\pi\ $sr in extent by measuring 
\v Cerenkov-light of shower particles in a water pond \citep{sulli-ic}.
A second significant advantage of MILAGRO is the high duty-cycle of
$\sim$ 100\%  compared with the $\sim$ 10\% duty cycle of sunlight and moonlight
limited observations with atmospheric \v Cerenkov telescopes.
The same is true for air shower arrays such as the TIBET air shower array 
\citep{amen-ic} or
ARGO-YBJ \citep{assi-ic}, which are composed of an array
of particle detectors to measure the secondary particles in
air showers produced by \grsn.

The number and quality of upcoming \gr observatories offer good prospects
for the future. Not only that the individual experiments will be much more
sensitive than their predecessors, also the energy range of satellite-based
and ground-based observatories will overlap, thus eventually providing 
complete coverage of the \gr spectra of sources from some 50 MeV to
10 TeV. The previously uncharted part of the spectrum, \gr energies between
10 GeV and 300 GeV, is particularly interesting, because leptonic \gr emission 
from supernova remnants should show a spectral peak in $\nu F_\nu$ 
representation, because competing models for \gr emission from pulsars 
predict different spectra, and because the infrared background light out to a
redshift of $z\approx 0.5$ should be measurable by virtue of absorption effects
in AGN \gr spectra.

\section{Diffuse galactic emission}

Why is it interesting to study diffuse galactic \grsn? This emission is produced
in interactions of cosmic rays with gas and ambient photon fields and thus
provides us with an indirect measurement of cosmic rays in various locations
in the Galaxy. A significant fraction of the diffuse galactic \grs is 
supposedly produced in decays of neutral pions following inelastic 
collisions of cosmic ray nucleons. Leptonic emission is particularly important
at \gr energies below 100 MeV, where bremsstrahlung is presumably the main 
emission mechanism. Inverse Compton scattering of relativistic electrons on 
soft ambient photons is expected to provide \grs with a hard spectrum,
thus eventually dominating over the $\pi^0$-decay \grs at high energies
\citep{port97}.
Measurements of diffuse galactic TeV \grs therefore constrain the
cosmic ray electron spectrum at multi-TeV energies.

Recent observations
made with the EGRET instrument on the Compton Gamma-Ray Observatory
of the diffuse Galactic \gr emission reveal a spectrum which is 
incompatible with the assumption that the cosmic ray spectra measured
locally hold throughout the Galaxy \citep{hu97}. The spectrum
observed with EGRET below 1 GeV is in accord with, and supports, the assumption
that the cosmic ray spectra and the electron-to-proton ratio
observed locally are uniform, however, the spectrum above 1 GeV, 
where the emission is supposedly dominated by $\pi^0$-decay, is harder
than that derived from the local cosmic ray proton spectrum. This is the 
well-known GeV excess.

\begin{figure} [tb]
{\includegraphics[width=8.5cm]{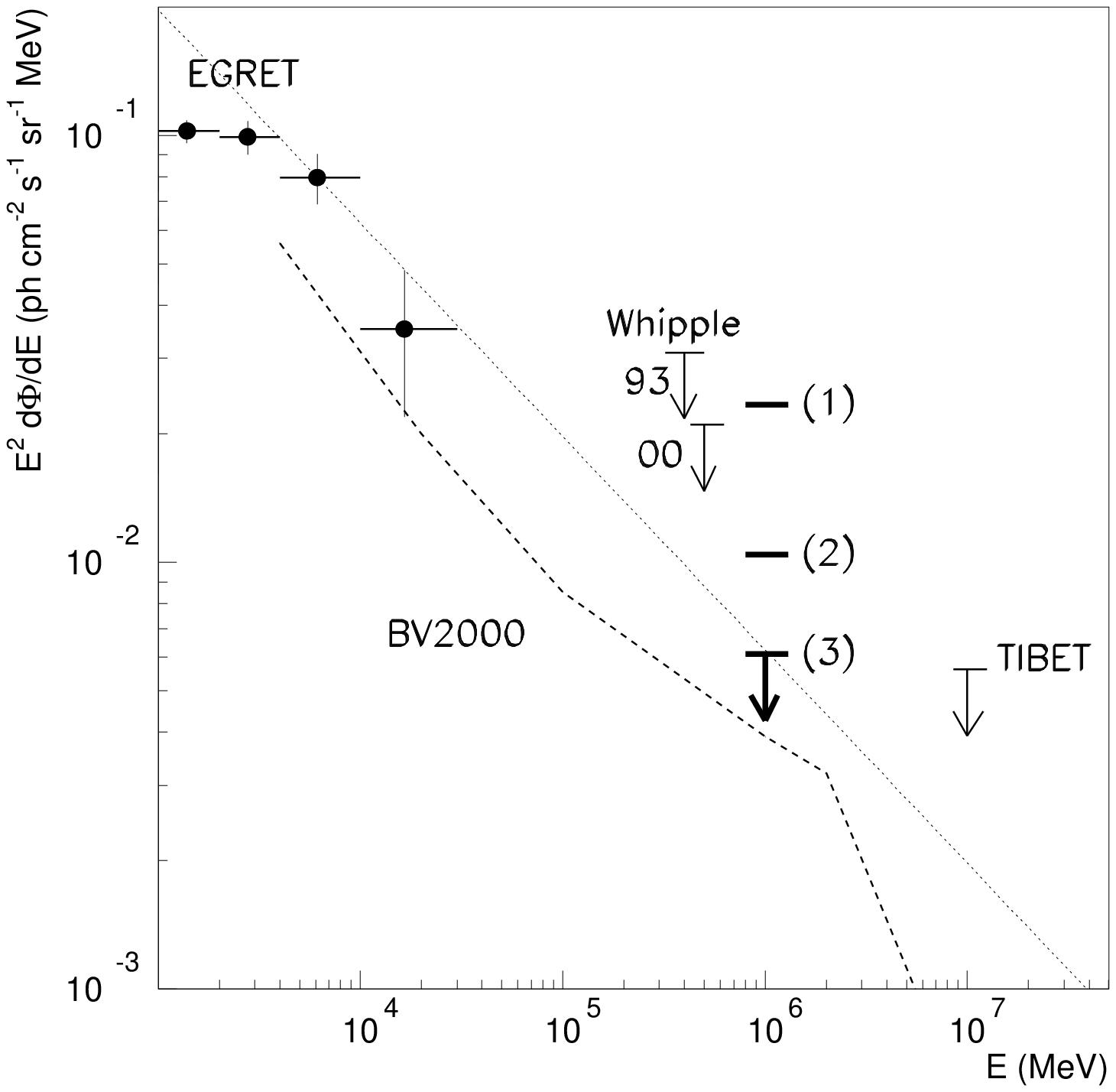}}
\caption{Upper limits for the diffuse Galactic \gr intensity derived by various
TeV \gr observatories (taken from \citet{lamp-ic}). The HEGRA upper limits
for the region $38^\circ < l< 43^\circ$, $\vert b\vert \le 2^\circ$
vary depending on what fraction of \grs is assumed to be due to diffuse Galactic emission \citep{ahar01}. Label 1 refers to all \grsn, label 2 to an 
independent data set for 
background subtraction, and label 3 to $\vert b\vert\ge 2^\circ$ data as background estimate. Also shown is the EGRET flux for 
$35^\circ < l< 45^\circ$, $\vert b\vert \le 2^\circ$, WHIPPLE upper limits
for $38.5^\circ < l< 41.5^\circ$, $\vert b\vert \le 
2^\circ$ \citep{rey93,lebo00}, and the TIBET upper limits \citep{ameno97}. 
The upper limits weakly depend on the spectral index in the 
respective energy range.
The dotted 
line is an extrapolation of the EGRET spectrum with an index of 2.5. The
dashed line refers to a model calculation by \citet{bere00} that 
is explained in the text.}
\label{lampeitl}
\end{figure}

\subsection{The GeV excess}

An interesting question is whether or not the GeV excess extends to TeV 
energies. WHIPPLE and HEGRA have observed a small field in the
Galactic plane at $l\approx 40^\circ$. The resulting spectrum of the
diffuse \gr emission in that field is shown in Fig.\ref{lampeitl}.
An upper limit at 10 TeV from the TIBET array is also shown, which however 
can not directly be compared with the results from the imaging \v Cerenkov
telescopes, for it represents the \gr flux from a much larger part of the sky.
The same restriction applies to new results from the TIBET II and 
TIBET HD arrays presented at this conference \citep{ameno-ic}.

The upper limits for diffuse Galactic \gr emission in the TeV energy
range slightly depend on the spectral index in the 
respective energy range. They also depend on what fraction of observed
\grs is attributed to the Galactic emission. Nevertheless it appears that
the spectrum of diffuse Galactic radiation between a few GeV and a TeV
can not be harder than a power law with a photon index of $\sim$2.4, which 
constrains published models of the GeV excess.

\citet{pe98} have argued that the local cosmic ray electron spectra
would not be representative for the Galaxy, if the electron were solely accelerated in supernova remnants (SNR). If the average electron spectrum 
in the Ga\-laxy is harder
than that measured locally, then the correspondingly hard spectrum of the
inverse Compton component could explain the GeV excess. The same basic
conclusion was drawn in a later study by \citet{smr00}. In these models
the inverse Compton spectrum is harder than $E^{-2}$ at a few GeV and displays
a slow softening at higher energies arising from the transition from
the Thomson regime to the Klein-Nishina regime for infrared target photons.
The inverse Compton spectrum would therefore violate the upper limits from
WHIPPLE, HEGRA, and TIBET, if the SNR as the assumed sources of cosmic ray 
electrons would produce single power law particle spectra extending
to electron energies higher than about 10 TeV. The available evidence for 
electrons with energies of 10 TeV or higher in SNR comes from observations
of non-thermal X-ray continua which are interpreted as synchrotron
radiation \citep{koy95}. It is interesting to note that for all SNR the
observed non-thermal X-ray flux is below extrapolations of the
radio synchrotron spectrum \citep{keoha97}, which indicates that SNR do not
produce electrons with single power law particle spectra extending
to electron energies higher than about 10 TeV. The upper limits for diffuse
galactic TeV \grs are therefore not in conflict with hard inverse Compton
models of the GeV excess.

It is possible that cosmic ray nucleons also contribute to the GeV excess.
\citet{bere00} have calculated the \gr yield of cosmic rays before escape
from their sources. The \gr spectrum produced within the sources would be
harder that of truly diffuse galactic \grs and thus unresolved sources of
cosmic rays should contribute significantly at TeV energies, but would 
presumably not explain the GeV excess (see Fig.\ref{lampeitl}). 

\citet{buesch01,buesch-ic} have investigated a dispersion in the 
cosmic ray
source spectra such that the SNR would produce power-law spectra with varying 
indices. Then the interstellar cosmic ray spectrum should 
display a curvature which could explain the GeV excess, provided the spectral
dispersion in the sources is sufficiently strong. Speculative
though B\"uschings model may appear, the radio spectra of SNR 
indicate that a spectral dispersion exists \citep{green00}, if  
somewhat smaller than required to explain the GeV excess in total. 
If his 
model was right, then the upper limits for the TeV \gr intensity would
require cosmic ray source spectra modified or cut off at about 100 TeV.

\subsection{Low energy \grsn}

\citet{dogiel-ic} have discussed the hard X-ray and soft \gr emission from the
Galactic ridge, which, if interpreted as diffuse emission and not caused
by unresolved sources, indicates the presence of a substantial flux of 
low energy cosmic rays. These authors find electron bremsstrahlung more likely 
than proton bremsstrahlung as the main radiation mechanism. More likely
though an electron origin appears, the required cosmic ray electron
source power would exceed the kinetic
power provided by supernovae and OB stars, suggesting that the 10 keV --
200 keV continuum emission from the Galactic ridge is still far from being
understood.

\section{Galactic sources}

\subsection{Supernova remnants}

SNR are considered the most likely sources of galactic cosmic rays, either
as individual accelerators or by their collective effect in superbubbles
\citep{bykov-ic}. Observational evidence in favor of this scenario has been
found only for cosmic ray electrons, not for the nucleons. 

\begin{figure*} [tb]
{\includegraphics[width=17cm]{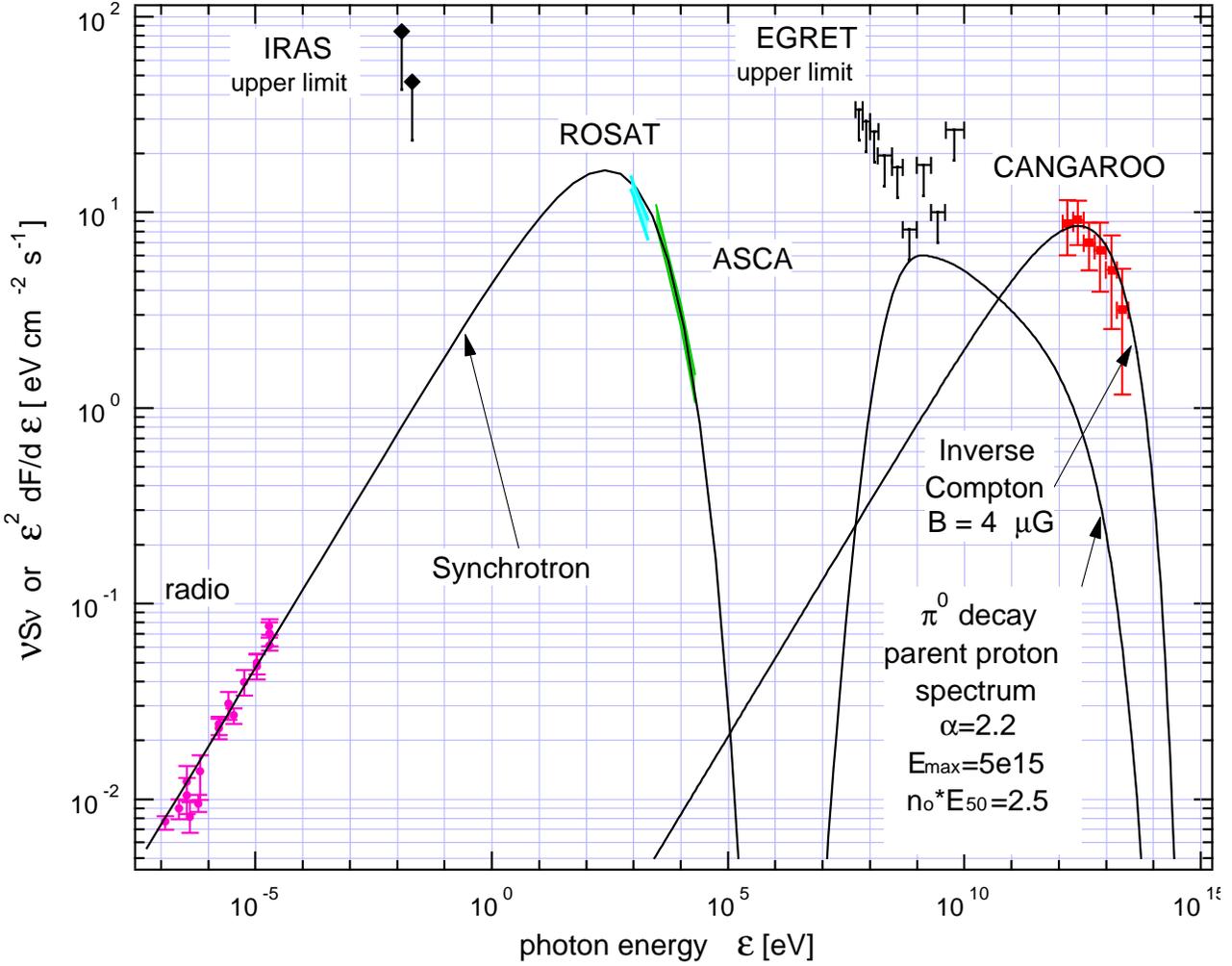}}
\caption{Multi-band spectrum of energy flux observed from the north-eastern
rim of SN~1006, where observed fluxes or upper limits of radio \citep{reyn98},
infrared, soft X-ray \citep{willi96}, hard X-ray \citep{ozaki98}, GeV \grsn,
and TeV \grs are presented (taken from \citet{tani-ic}). The solid lines 
represent their fits based on 
an inverse Compton model and $\pi^0$-decay.}
\label{tani-fig}
\end{figure*}

Three shell-type SNR have been detected at TeV \gr energies so far. SN~1006
has been reobserved with CANGAROO with a flux consistent with the previously 
published result \citep{hara-ic}. Also in the new observations only the north-eastern rim is seen in TeV \grsn. Recent observations of RX~J1713-3946
with the 10m-telescope CANGAROO-II have yielded a detection with about
$8\sigma$ significance and have thus
confirmed the earlier measurement \citep{enomoto-ic}.
The HEGRA array of \v Cerenkov telescope has detected TeV \grs from 
Cassiopeia~A \citep{puehl-ic}, if with 0.03~Crab above 1 TeV
at a flux much lower than 
those reported for the two southern remnants.

All three shell-type
SNR detected so far show non-thermal X-ray emission, which presumably
is synchrotron radiation. It is known that the synchrotron radiating 
electrons would inverse-Compton scatter the microwave background to TeV \gr
energies with a flux depending only on the X-ray flux and
the magnetic field strength within the remnant \citep{pohl96},
provided both are measured at photon energies corresponding to the same
electron energy. For the two 
southern remnants SN~1006 and RX~J1713-3946 a significant contribution of
\grs from hadronic interactions appears unlikely on account of the low density
environment in which the remnants reside. 

\subsubsection{SN~1006 and RX~J1713-3946}

We have already noted that for all SNR the
observed non-thermal X-ray flux is below the extrapolation of the
radio synchrotron spectrum \citep{keoha97}, implying a cut-off in the cosmic ray
electron spectrum. The actual cut-off energy would depend on the 
magnetic field strength, for it is measured in synchrotron frequency. The 
interesting question of whether or not the cut-off would be caused by energy
losses, implying whether or not a similar cut-off must be expected in the
cosmic ray nucleon spectra, is also a question of the magnetic field strength,
for synchrotron radiation is the main energy loss channel.

Two important issues need to addressed:

-- Are the X-ray and TeV \gr spectra of SN~1006 and RX~J1713-3946 compatible with each other in the sense of both being produced by the same particles? If
that was the case, it would confirm our notion of an inverse Compton origin 
of the \grs and we could indeed use the TeV flux as a measure of the magnetic 
field strength.

-- Is the magnetic field strength thus determined such that the high energy
cut-off in the electron spectra can be caused by synchrotron energy losses? If
that was not the case, the cause of the cut-off would have to be intrinsic to
the actual acceleration process and therefore also affect the cosmic ray 
nucleon spectra, which then would not be single power laws up
to the knee at a few PeV.

\citet{tani-ic} find the \gr spectrum of SN~1006 between 1.5 TeV and 20 TeV
well described by a power law
\begin{displaymath}
J(E) =(1.1\pm 0.4)\,10^{-11}\,\left({E\over {\rm TeV}}\right)^{2.3\pm 0.2}
\end{displaymath}
\begin{equation}
\hphantom{J(E) =(1.1\pm 0.4)\,10^{-11}}
{\rm TeV^{-1}\,cm^{-2}\,sec^{-1}}
\label{1006spec}
\end{equation}
When assuming a power law with exponential cut-off for the electron spectrum
a fit of the combined radio, X-ray and \gr data is obtained
with a cut-off energy $E_{\rm c} \simeq 50\ $TeV and a
magnetic field strength $B\simeq 4\ \mu$G (see Fig.\ref{tani-fig}).
\citet{allen-ic} have carefully analyzed the X-ray spectrum between 0.12 keV and
17 keV and find that the best model includes a thermal component and
a broken power law component
($s_1=2.08\pm 0.14$, $s_2=3.02\pm 0.17$, and the break energy $E_c
=1.85\pm 0.2\ $keV) to describe the non-thermal continuum. Given the best-fit
$B\simeq 4\ \mu$G of \citet{tani-ic}, a \gr energy of 5 TeV would correspond to
an X-ray energy of 0.4 keV. Based on their findings for the non-thermal part
of the X-ray spectrum and the earlier \gr measurements \citep{tani98},
\citet{allen-ic} have also presented a fit to the multi-band spectrum of 
SN~1006. With their parameters $B\simeq 10\ \mu$G and 
$E_{\rm c} \simeq 20\ $TeV a \gr energy of 5 TeV would correspond to
an X-ray energy of 1 keV. The \gr spectrum measured with CANGAROO is statistically well defined below 10 TeV and thus 
has to be compared to the low energy X-ray spectrum with which it agrees.
The compatibility of the X-ray and \gr spectra of SN~1006 supports
our notion of an inverse Compton origin of the \grsn. A confirmation
would require the observation of corresponding curvature in the X-ray
and TeV \gr spectra, though.

The magnetic field strength of $B\simeq 4\ \mu$G found by \citet{tani-ic}
is disturbingly low, for the magnetic field in the rim should be compressed.
The local upstream field around SN~1006 would have to be 
$B_{\rm up}=1-2\ \mu$G depending on orientation. We can calculate the 
e-folding acceleration time for diffusive shock acceleration with the
diffusion coefficient $D=\eta D_{\rm Bohm}$, $\eta \ge 1$, written in 
units of the Bohm diffusion coefficient $D_{\rm Bohm} = c r_g /3$ with $r_g$ 
as the Larmor radius of the electrons.
Then
\begin{displaymath}
\tau_{\rm acc} \simeq {{4\,D}\over {v_{\rm shock}^2}}
\end{displaymath}
\begin{equation}
\hphantom{\tau_{\rm acc}}
\simeq (600\ {\rm years})\ \eta\ {{
\left({E\over {\rm 50\ TeV}}\right)}\over {
\left({B\over {\rm 4\ \mu G}}\right)
\left({{v_{\rm shock}}\over {\rm 3000\ km/sec}}\right)^{2}}}
\label{tauacc}
\end{equation}
The acceleration time is similar to the age of the remnant for Bohm diffusion,
i.e. $\eta=1$. In the general case $\eta \gg 1$ the diffusive shock acceleration
would not operate sufficiently rapid to provide electrons with 50 TeV within
the age of SN~1006. If the magnetic field strength was substantially higher
than $4\ \mu$G, the acceleration time would be correspondingly smaller. As
a result electron acceleration to 50 TeV within the time given  would appear
more feasible.

\citet{swaluw-ic} have combined hydrodynamical calculations of the evolution
of a young shell-type SNR with an algorithm, which simultaneously calculates
the associated particle acceleration in the test-particle approximation.
These authors have not modelled the TeV spectrum
in parallel to the X-ray spectrum. Nevertheless, they find that at an age of 
1000 years a substantial fraction of accelerated electrons would have escaped from the regions of compressed magnetic field in the rims of SN~1006. While all
electrons would comptonize the microwave background to TeV energies, only 
a fraction of them would emit synchrotron radiation in a high magnetic field
region. It is actually possible to obtain a fit to the multi-band spectrum
of SN~1006 by assuming that the magnetic field occupies only 40\% of the volume
filled with cosmic ray electrons \citep{allen01}. In this case the 
best-fit parameters
would be $B\simeq 40\ \mu$G and $E_{\rm c} \simeq 10\ $TeV. The electron
energy loss time scale for synchrotron radiation at the energy $E_{\rm c}$
would be 900 years and thus similar to the age of the remnant. The acceleration 
time (Eq.\ref{tauacc}) would be similar or smaller than both the age and the
energy loss time for a diffusion coefficient
\begin{displaymath}
\tau_{\rm acc}\le \tau_{\rm loss}\simeq \tau_{\rm age}
\end{displaymath}
\begin{equation}
\Rightarrow\qquad \eta \le 75 
\left({{v_{\rm shock}}\over {\rm 3000\ km/sec}}\right)^{2}
\label{eta}
\end{equation}
which would comfortably allow diffusive shock acceleration to accelerate
electrons to the energies observed for a fair range of intensities of electromagnetic turbulence. \citet{allen01} have assumed the
extreme case of a vanishing magnetic field in part of the volume. A realistic 
scenario would foresee a compressed magnetic field in the rims of the
remnant and a lower (by a factor of a few) magnetic field strength outside the
rims of SN~1006. Then we can expect 
$B\approx 20\ \mu$G in the rims, $E_{\rm c} \approx 15\ $TeV, and 
$\tau_{\rm loss}> \tau_{\rm age}$.

We can therefore conclude that the electron spectrum in SN~1006 is probably not 
significantly modified by energy losses on account of the energy loss 
time 
being similar or larger than the age of the remnant. If cosmic ray nucleons 
were accelerated in parallel
to the electrons, their spectrum would presumably show the same cut-off
energy $E_{\rm c}$ as does the electron spectrum. Is it possible that during the
later evolution of SN~1006 nucleons are accelerated to the knee at a 1000 TeV?

In the standard hydrodynamical model of SNR their evolution has a first phase,
in which the expansion proceeds with constant velocity, followed by the
so-called Sedov phase, during which the outer shock decelerates. The 
deceleration of the shock causes diffusive shock acceleration to operate
less efficiently (see Eq.\ref{tauacc}), so that the maximum particle energy
can increase only by a factor of a few during the Sedov phase. The question
whether or not SN~1006 can accelerate cosmic ray nucleons to the knee is
therefore linked to the question whether or not SN~1006 is already in the
Sedov phase; a question to which I can not give a firm answer.

\subsubsection{Cassiopeia A and Tycho}

The supernova remnant Cassiopeia A differs from SN~1006 and RX~J1713-3946
in that the supernova blast wave is expanding into a wind bubble and
shell system from the previous wind phases of the progenitor star
\citep{bork96}. The matter density and the magnetic field strength in
the upstream region of the outer shock are those of a red supergiant wind
and not those commonly found in the interstellar medium. 
Cas~A also shows a non-thermal hard X-ray continuum \citep{allen97},
which would imply high energy \gr emission from inverse Compton scattering.
However, we must expect that both the \gr flux and the cut-off energy in the 
\gr spectrum are much less than for SN~1006 and RX~J1713-3946 on account
of the much higher magnetic field strength in Cas~A, for which
estimates for the magnetic field strength at the shock and in the
downstream region suggest $B_{\rm d} \simeq 1\ $mG \citep{atoyan00}.

\begin{figure} [tb]
{\includegraphics[width=8.5cm]{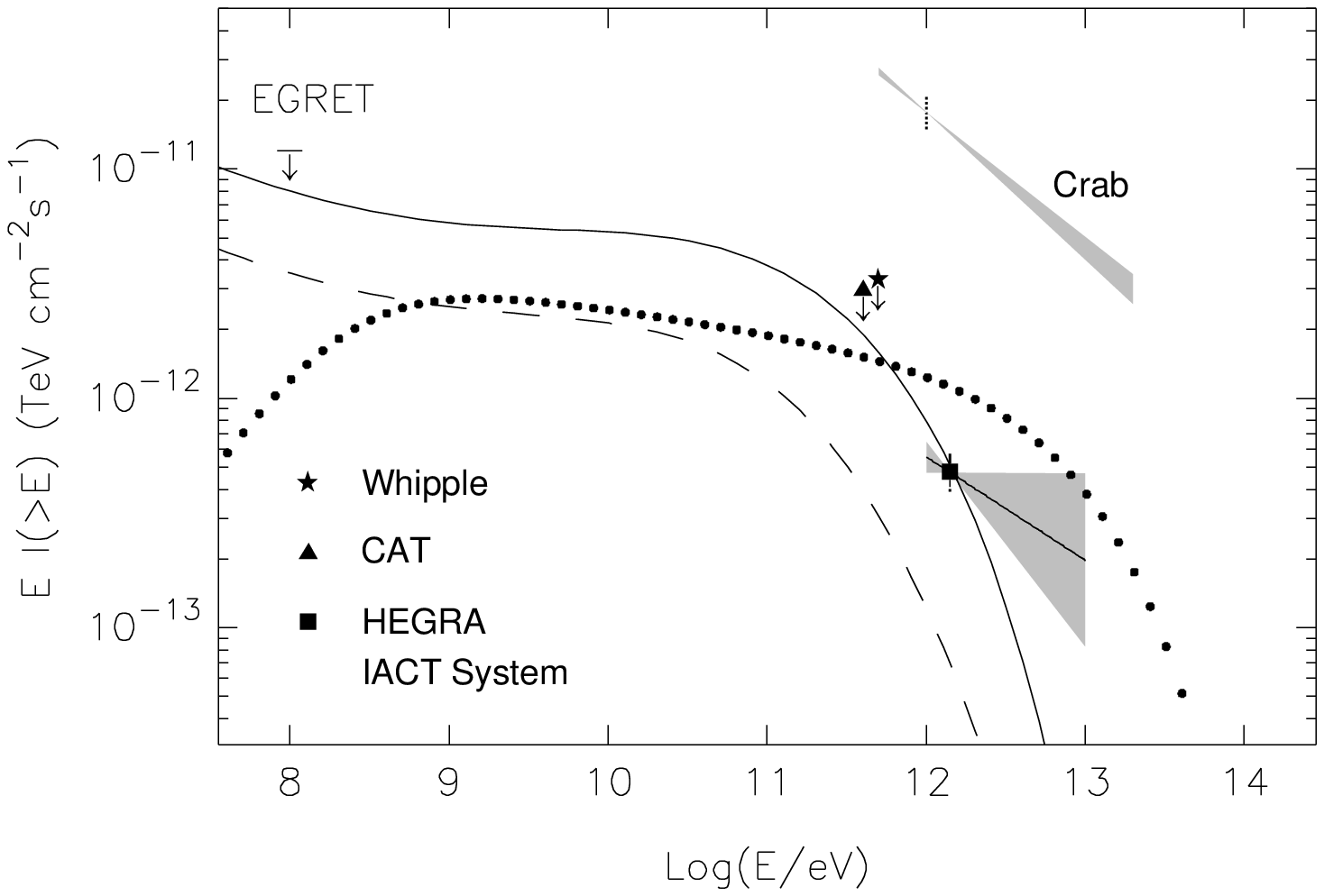}}
\caption{The measured TeV \gr flux and spectral index of Cas~A
in the context of model predictions (taken from \citet{puehl-ic}).
The shaded area shows the 1$\sigma$ error range for the measured spectral
distribution under the assumption of a power law spectrum.
Also indicated are the upper limits measured by EGRET, WHIPPLE
\citep{less99}, and CAT \citep{goret99}. The data are compared with 
model predictions published by \citet{ato00}.
The dotted curve represents a model spectrum for the \gr flux arising from 
$\pi^0$-decay. The solid and the dashed line show the predicted
inverse Compton plus bremsstrahlung spectra for two different parameter sets.}
\label{casa-fig}
\end{figure}

The measured \gr flux and spectrum of Cas~A are shown in Fig.\ref{casa-fig}. 
Because of the moderate statistical significance of the overall detection,
the spectral index is only poorly constrained. Also shown in the figure 
are model spectra based on calculations by \citet{ato00}. To be noted
from the figure is that the predicted flux from $\pi^0$-decay exceeds
the observed flux (the prediction was made prior to the actual detection).
The problem is that the expected absolute flux level of $\pi^0$-decay
\grs is not well determined in the context of general acceleration
models. One of the crucial but poorly known parameters is the 
injection efficiency, with which suprathermal protons are injected at 
the shock front. In contrast to the electrons, for which the non-thermal
X-ray flux can be used as a primer for the electron flux, whatever the
micro-physics at the acceleration site, the high energy cosmic ray 
nucleons do not reveal themselves in any observable channel other than
\gr emission.

The injection efficiency does affect the overall efficiency of
SNR in transferring their bulk kinetic energy to a few high energy cosmic rays.
A high injection efficiency, as assumed by \citet{ato00}, would provide
SNR with sufficient cosmic ray source power to constantly replenish the
galactic cosmic rays. The data of Cas~A suggest that some crucial parameters
of the acceleration process are actually less favorable than assumed in the
theoretical studies. To date we can not say, whether or not the TeV \gr data
are in conflict with our notion that SNR accelerate the bulk of cosmic ray 
nucleons to PeV energies, but the situation is getting tight.

A high cosmic ray density in the remnants also implies backreactions
of the cosmic rays on the acceleration process, one of which is a modification
of the shock compression ratio caused by the pressure and energy density of the 
cosmic rays. \citet{berezhko-ic} have presented a calculation of
non-linear particle acceleration in Cas~A. They have determined the
injection rate for electrons by a fit of the radio to X-ray spectrum.
The proton injection required to fit the observed TeV \gr spectrum
of Cas~A would be more than an order of magnitude less than the
electron injection rate at the same energy. It has been suggested
that in a quasi-perpendicular shock electrons can be efficiently
injected, whereas proton injection is suppressed \citep{md01}. On
the other hand, turbulence in the progenitor wind and
at the interface with swept-up material should lead to many
field lines being locally shock-parallel, thus allowing efficient proton 
injection at some parts of the shock. Clearly, more studies of the 
microphysics of particle injection at the shocks are needed.

\begin{figure} [tb]
{\includegraphics[width=8.5cm]{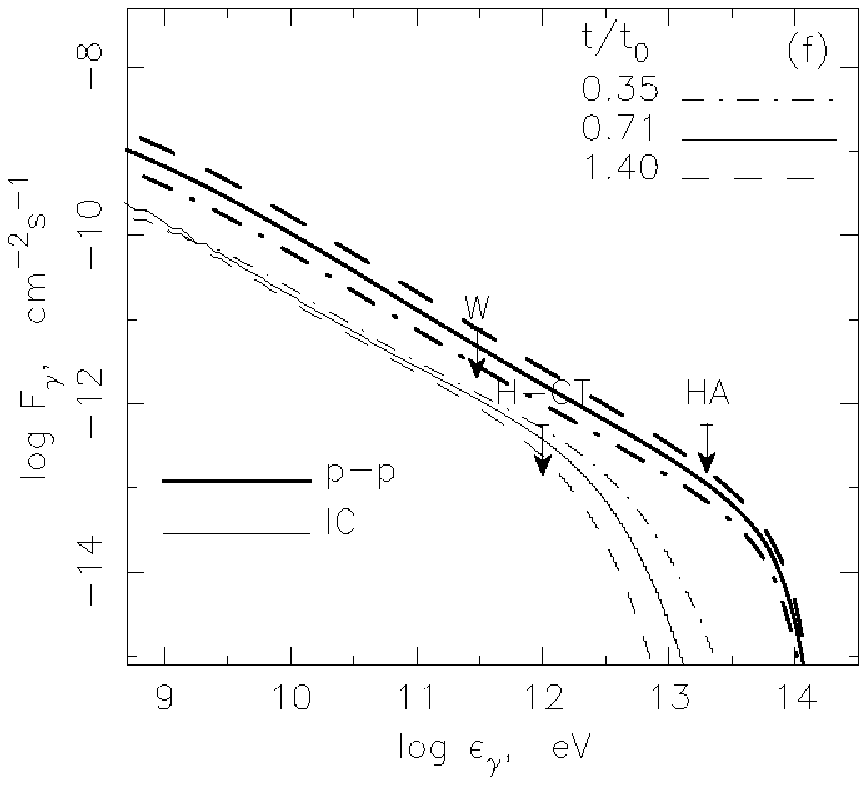}}
\caption{The upper limits for TeV \grs from Tycho's SNR in comparison
with the temporal evolution of \gr emission according to the model
of \citet{voelk-ic}, from where the figure has been taken. The flux 
limits are W: WHIPPLE \citep{buck98}, H-CT: HEGRA IACT system
\citep{aharo01}, and HA: HEGRA AIROBICC \citep{prahl97}.
The modelled \gr spectra are shown for three evolutionary phases
with the solid lines corresponding to the current stage of Tycho's evolution.
The time $t_0 $ marks the turn-over into the Sedov phase.}
\label{tycho-fig}
\end{figure}

No \gr emission has been detected from Tycho's SNR so far.
HEGRA has established a very low 3$\sigma$ upper limit of 0.03 Crab
above 1 TeV \citep{aharo01}. In X-rays Tycho shows a thermal spectrum
with strong lines and a bremsstrahlung continuum, but also a hard X-ray 
tail \citep{petre99}, which presumably is of non-thermal origin. If we
interpret the apparently non-thermal X-ray emission as synchrotron radiation,
the upper limit for the TeV \gr flux implies a lower limit for the magnetic 
field strength with $B\ge 20 \mu$G, when the hard X-ray tail is extrapolated
to lower X-ray energies, or $B\ge 6 \mu$G, when the non-thermal X-ray flux is
estimated by modelling ASCA data \citep{hwang98}.

\citet{voelk-ic} have used a nonlinear kinetic model of cosmic ray acceleration similar to the one applied to Cas~A \citep{berezhko-ic}
to describe the properties of Tycho's SNR and to model the \gr emission.
These authors argue that a magnetic field strength $B_0 = 40\ \mu$G in the upstream region and consequently $B_d \approx 200\ \mu$G in the downstream
region is required to reproduce the synchrotron spectrum from radio
to X-ray frequencies. In such strong magnetic fields
high energy electrons in the downstream region are subject to radiative 
energy losses on a time scale much shorter than the age of remnant.
Therefore the electron spectrum displays a turnover to a softer
power law at about one TeV with a spectral index change $\Delta s =1$.
The spectral indices of the TeV scale
\gr spectra of inverse Compton scattering and 
$\pi^0$-decay are then similar, however, the inverse Compton spectrum
cuts off at a much smaller energy than does the $\pi^0$-decay spectrum,
as shown in Fig.\ref{tycho-fig}. As in case of Cas~A the expected 
hadronic \gr flux exceeds the observed value or limit, if for Tycho only by
a factor of ten. Presumably the cause of that discrepancy is the same in 
both cases. The overprediction of the hadronic TeV \gr flux from 
Cas~A and Tycho also compromises corresponding model predictions for hadronic 
\gr emission from SN~1006 \citep{bere-ic}.

\subsection{Unidentified EGRET sources}

EGRET has left a legacy of about 170 sources not yet identified firmly with
known sources. Various population studies have been performed to search 
for correlations with classes of galactic objects. It has been noted only 
recently that very much care has to be exercised in these studies to 
account for systematic effects arising from the uneven exposure 
distribution and the structured galactic foreground emission \citep{reito-ic}.
Searches for TeV \grs in the EGRET error boxes have not been successful so far
\citep{fega-ic}.

It has been suggested that some of the unidentified EGRET sources are
SNR \citep{espo96}. A careful study shows that the spectra of well
observed SNR candidates, associated with CTA 1, W28, IC443, and $\gamma$ Cygni,
are suggestive of a pulsar origin rather than young cosmic rays in shell-type
SNR \citep{reibe-ic}. It is in fact possible that a number of 
unidentified \gr sources are actually pulsars born in the local star-forming 
region Gould's belt \citep{perrot-ic}.

\subsection{Pulsars and plerions}

To date eight pulsars have been identified in the EGRET data 
on account of pulsed
emission. There are two competing models for the production of pulsed  \grsn:
the polar cap model \citep{hard99} and the outer gap model \citep{hiro01}, which may be observationally distinguished in the energy range
between 3 GeV and 30 GeV. The extent of pulsed emission to very high
\gr energies is a unique prediction of the outer gap models,
and is not permitted by polar cap models.
The non-imaging \v Cerenkov telescopes
STACEE \citep{oser00} and CELESTE \citep{dumora-ic} have now established
upper limit for the pulsed flux of the Crab at \gr energies of 190 GeV and
60 GeV, respectively. In their final configurations these two experiments
will operate with substantially lower energy thresholds, as will do 
MAGIC, and thus
will allow observational tests of the outer gap models.

There are a handful of pulsar-powered SNR with syn\-chro\-tron nebula, so-called
plerions. The Crab nebula is the prototype plerion and serves as a
standard candle in high energy astrophysics. Unpulsed \gr emission, which is
commonly interpreted as being caused by inverse Compton scattering, can be 
detected up to about 20 TeV \citep{aha00, ameno00}. 

In case of PSR 1706-44/G343.1-2.3
the energy density of the synchrotron radiation in the nebula is less than 
that of the microwave background. Because the latter is known, quantitative
estimates of the unpulsed TeV emission can be made \citep{aha97}. Earlier measurements of PSR 1706-44 with CANGAROO \citep{kifune95} and the Durham Mk6
telescope \citep{chad98} have indicated a TeV flux, which is an order of
magnitude higher than predicted. Recent measurements with CANGAROO II 
have confirmed the high flux level and spectrum \citep{kush-ic},
so that our problem to understand the source persists.

\section{Active galactic nuclei}

\begin{figure} [tb]
{\includegraphics[width=8.5cm]{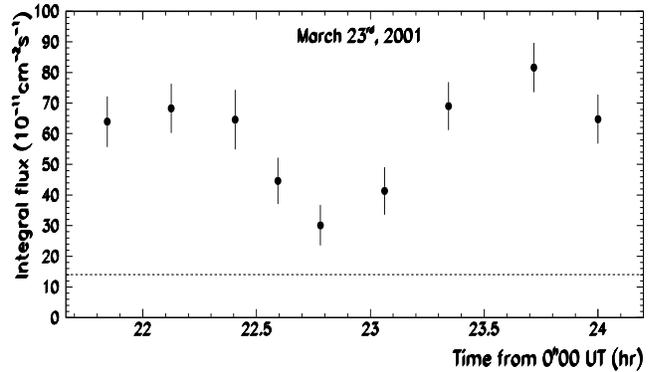}}
\caption{Mkn 421 integral flux above 250 GeV during the night of 
March 23rd, 2001, as observed with the CAT telescope (taken from 
\citet{khel-ic}). Each point indicates a $\sim$ 15 min observation, the dashed
line shows the flux level of the Crab nebula. The \gr flux from Mkn 421
varies by more than 2 Crab within one hour.}
\label{khelifi-fig}
\end{figure}

Why is it interesting to study \grs from active galactic nuclei (AGN)?
These sources show very intense emission, which in many cases is variable.
The variability has been observed on all time scales accessible with 
the available measurement techniques down to about one hour (see 
Fig.\ref{khelifi-fig}). It should be noted that the AGN detected
in the GeV to TeV range emit a significant, if not dominant, fraction of their
luminosity in the form of \grsn, indicating that with measuring \grs we
actually study the main energy transfer processes in these objects.

A many studies suggest a flux correlation between X-ray and TeV \gr emission of
AGN, which, if real, would allow a complementary view of the radiating
particles, whatever their nature.

TeV \gr astronomy also provides means to probe the intergalactic infrared
background radiation by measuring the absorption due to photon-photon
pair production. 

The types of AGN detected at high energies, which include flat-spectrum radio
quasars (FSRQ) and BL Laceratae objects (BL Lacs), are collectively referred
to as {\it blazars}. The broadband emission from blazars from radio wavelengths
to the UV -- or even X-rays in some cases -- is apparently dominated
by highly beamed, incoherent synchrotron radiation produced in a relativistic
jet aligned closely to the direction to the observer. The relativistic
beaming results in a strong amplification of the apparent luminosity and a
reduction of the apparent variability time scales. It also explains the
frequent observations of superluminal motion in these sources.
In blazars \gr observations reveal a second component of the spectrum
which does not connect smoothly with the low energy component.
The multiband spectrum of blazars thus has a double hump shape with the
second component peaking at energies between a few MeV and a few TeV.
The underlying radiation mechanism of the high energy component is still
the subject of debate, as is the nature of the particles causing the radiation.
Nevertheless it has been recognized that inverse Compton scattering by the
synchrotron radiating electrons should contribute to the \gr component
in the multiband spectra of blazars. 

EGRET has detected about 70 blazars in the energy range between 100 MeV
and 10 GeV, most of which are FSRQ. The imaging atmospheric \v Cerenkov
telescopes have observed a handful of blazars in the TeV energy range, 
most of which are BL Lacs. BL Lacs are noted for a very small contribution of 
thermal emission in the Optical on acount of the small equivalent width
of lines in the spectrum. It is possible that in BL Lacs the ambient soft photon
field is much more dilute than in FSRQ, implying that internal absorption by
pair production would be less efficient and that inverse Compton scattering of 
these photons would not play an important r\^ ole, both compared with FSRQ and 
with inverse Compton scattering of synchrotron photons produced in the jet,
the so-called synchrotron-self-Compton (SSC) process. It appears that
soft photons from a possibly existing dust torus are particularily important
in FSRQ \citep{donea-ic}, depending on the geometrical structure of the
torus \citep{arbeiter-ic}.

The threshold energy for pair production is identical to the energy at which
the transition between the Thomson regime and the Klein-Nishina regime in 
inverse Compton scattering occurs. \citet{georg-ic} have extended earlier
treatments of the problem \citep{mause} and have shown that the transition to 
the Klein-Nishina regime causes a turn-over to softer \gr spectra at higher
energies, which can seriously compromise interpretations of the spectral energy
distribution in the framework of an SSC origin of the high energy radiation.

Very little is known on the origin of the radiating particles in the jets of 
AGN. Are the synchrotron radiating electrons the primary particles, i.e. 
directly accelerated in the jets, or are they secondary particles produced in 
inelastic interactions of high energy nucleons? In the latter case neutrinos
would be produced in parallel to the \grsn. \citet{schus-ic} have calculated
the neutrino yield for a particular model of particle energization in the jet
and have found that the detection of neutrinos from AGN would be possible with
future neutrino telescopes of the ICECUBE class, if the \gr light curves of blazars are used to define the search windows in data space.

\subsection{Which AGN have been observed in TeV \grsn?}

The BL Lacs Mkn 421 and Mkn 501 are now regularily observed by many groups.
Both sources are usually so bright that well defined \gr spectra can be 
obtained. Some other BL Lacs have been observed in the past, but haven't been 
detected in later campaigns or by other groups. An example is 2344+514 which has
been detected with WHIPPLE \citep{cata98}, but not with HEGRA, and has now
been detected again with WHIPPLE, if with very moderate statistical significance
\citep{badran-ic}. The southern BL Lac 2155-304 has not been 
detected with CANGAROO II in the year 2000 \citep{nish-ic} with a flux limit 
below the level previously reported \citep{chad99}. The prototype of the 
class, BL Lacertae (2200+420) has also not been detected in recent observations 
\citep{mang-ic} with a flux limit 
below the level previously reported \citep{nesh01}.
A good candidates though it is,
2005-489 \citep{nish-ic} was not detected as a source of TeV \grsn.

Recently the BL Lac 1426+428 was detected with WHIPPLE \citep{horan-ic} and
confirmed with HEGRA and CAT (in both cases communicated only
at the conference). This source is interesting for its redshift of 
z=0.129 which is about four times that of Mkn 421 and Mkn 501, thus allowing 
studies of the effect of absorption by the infrared background.

Not detected in TeV scale \grs at all to date are radiogalaxies and quasars
\citep{lebohec-ic,goett-ic}.

\subsection{Correlation between X-rays and \grsn}

Searches for possible correlations are a standard tool in astronomy when the
basic characteristics of sources have to be understood. The multiband
spectra of blazars typically show a low energy and a high energy component, which are possibly produced by the same particles through different radiation 
processes, e.g. synchrotron radiation at low energies and inverse Compton 
scattering at high energies. It may therefore be useful to compare the 
lightcurves of blazars at the energies at which the components display their 
peak in emitted power, namely X-rays and TeV \grsn.  

\begin{figure} [tb]
{\includegraphics[width=8.5cm]{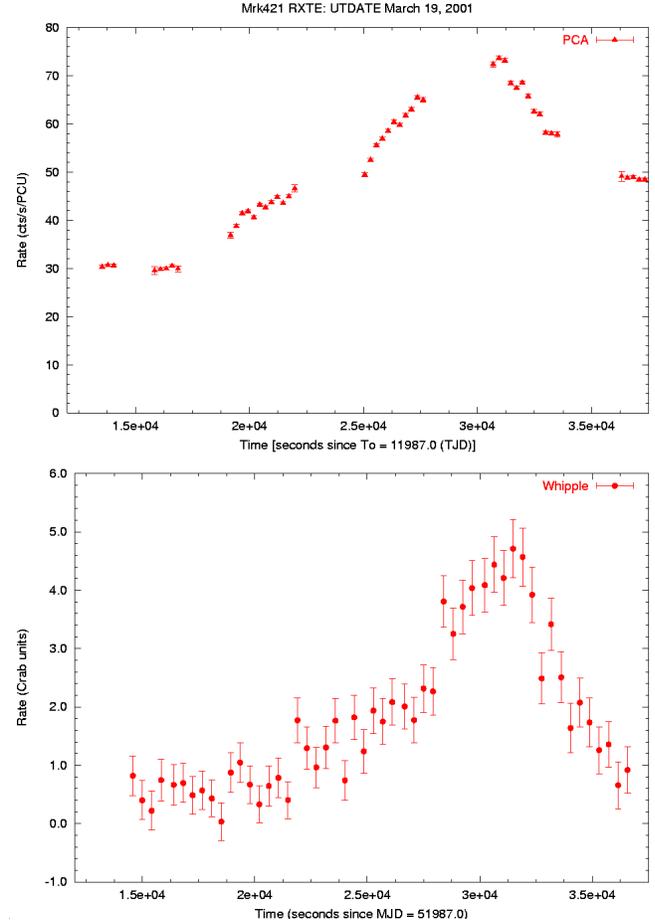}}
\caption{Simultaneous X-ray/\gr flare observed on March 19, 2001 (taken from 
\citet{jordan-ic}). The 2-10 keV X-ray light curve was obtained with the 
PCA detector on RXTE \citep{fossati01}. The E$>$300 GeV \gr data were obtained
with the WHIPPLE telescope and are binned into 4 min intervals.}
\label{jordan-fig}
\end{figure}

This can be done for short, but well covered periods of time, an example of which is shown in Fig.\ref{jordan-fig}. One particular outburst in TeV \grs 
happens to coincide with one outburst in X-rays without noticable delay.
The figure shows the rising phase and the decay phase of the outbursts, but not
the behaviour preceding or following the event. Apparently the X-ray and \gr light curves are well correlated for the particular interval of seven hours 
displayed here. Does that imply that we can speak of a X-ray/\gr correlation?
Or are we guilty of sample occulting by selectively showing the data
when the fluxes vary in unison?
I will come back to this point later.

What conclusion on the physics in the jet of Mkn 421
can be drawn given the rapid outburst displayed in Fig.\ref{jordan-fig}?
A number of authors have dealt with this subject in their presentations 
and I repeat the main arguments here.
The TeV scale \gr outburst has a rise time scale of about one hour and a
similar or possibly somewhat shorter decay time scale. The decay is
probably related to energy losses and thus to internal processes. Accounting
for relativistic beaming by the Doppler factor, $D$, we find the energy loss time scale in the jet frame as
\begin{equation}
\tau^\ast = D\,\tau_{\rm obs} \simeq 3000\ D\quad{\rm sec}
\label{1}
\end{equation}
The power emitted in X-rays is a significant fraction of the observed 
bolometric luminosity, therefore under the assumption of a synchrotron origin 
of the X-rays the electron energy losses can be approximated by those for
synchrotron radiation, yielding
\begin{equation}
\tau_{\rm sy} (10\ {\rm keV}) \simeq 1200\ \left({B\over {\rm Gauss}}\right)^{-1.5}\quad{\rm sec}
\label{2}
\end{equation}
so that
\begin{equation}
\tau_{\rm sy} (10\ {\rm keV}) \simeq \tau^\ast
\quad\Rightarrow\quad B\simeq 0.5\ D^{-2/3}\quad {\rm Gauss} 
\label{3}
\end{equation}
For information travel time reasons the source region has a diameter not larger than
\begin{equation}
R^\ast \le R_{\rm c}\approx  c\,\tau^\ast \simeq 10^{14}\ D\quad{\rm cm}
\label{r3}
\end{equation}
Given these relations we can ask two questions.
What are the source size and magnetic field strength required for the 
SSC emission not to exceed the observed TeV \gr flux?
Then, assuming the X-rays are produced in the same volume as are the \grsn,
what is the optical depth of \grs for pair production with the X-ray photons?

\begin{figure} [tb]
{\includegraphics[width=8.5cm]{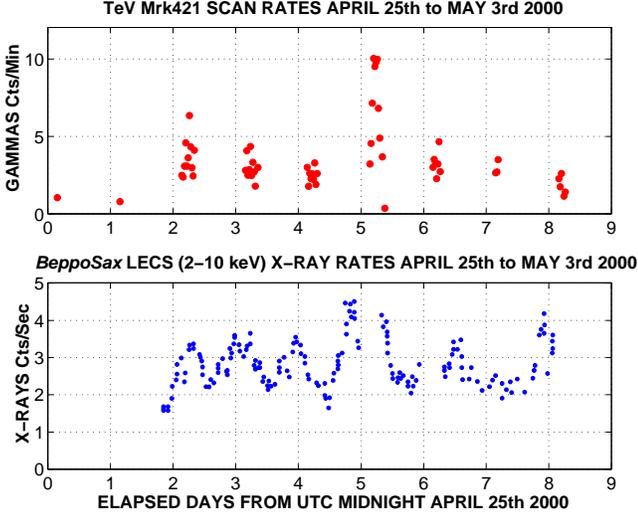}}
\caption{Contemporaneous Mkn 421 TeV \gr and X-ray light curves (taken from 
\citet{feg-ic}). The \gr data have been obtained with the WHIPPLE telescope.
The maxima in the count rates in \grs and in X-rays apparently coincide, though
the X-ray detector had a dropout at the time of the peak in \grsn. The 
variability in X-rays is fairly well resolved, whereas that in \grs is clearly 
undersampled.}
\label{fegan-fig}
\end{figure}

Let us first deal with the SSC question. In the jet frame the energies of the X-ray and \gr photons can be written in units of the electrons rest mass energy.
\begin{equation}
\epsilon_X^\ast \simeq {{0.01}\over D}\qquad \epsilon_\gamma^\ast \simeq
{{10^6}\over D}
\label{4}
\end{equation}
so that the comptonization of X-rays would happen in the Klein-Nishina regime
with a cross section reduced by a factor $\sim D^2\,10^{-4}$ compared with the Thomson
cross section, $\sigma_T$. 
Let us compare the energy density of X-ray photons in the source at the luminosity distance $r_L$ with that of the magnetic field.
\begin{equation}
u_{\rm ph}^\ast \simeq {{r_L^2}\over {c\,D^4\,{R^\ast}^2}}\,(\nu F_\nu)
\ge {{10^5}\over {D^6}}\quad{\rm erg\,cm^{-3}}
\label{5}
\end{equation}
This energy density weighted with the reduction factor for the scattering cross 
section $\sim D^2\,10^{-4}$ must not be larger than the magnetic energy 
density, otherwise the 
expected SSC \gr flux would exceed the observed flux. This leads to a lower 
limit for magnetic field strength, which together with Eq.\ref{3} gives a lower limit for the Doppler factor.
\begin{equation}
B\ge 15\ D^{-2}\quad {\rm Gauss}
\qquad\quad \Rightarrow \qquad D\ge 13
\label{7}
\end{equation}
Please note that $D$ could be smaller if the synchrotron energy loss time scale 
is smaller than the flare decay time scale (Eq.\ref{1}).
Eq.\ref{4} indicates that for Doppler factors $D\le 100$ the \gr photons can 
produce pairs by collisions with the X-ray photons with a cross section
of the order of $\sigma_p \simeq D^2 \sigma_T 10^{-4}$. The optical depth is 
approximately 
\begin{equation}
\tau_p \simeq {{u_{\rm ph}^\ast\, R^\ast\,\sigma_p}\over 
{\epsilon_X^\ast \,m_e c^2}} \ge 0.1 \ D^{-2}
\label{8}
\end{equation}
so that internal absorption on the X-ray photons is not important,
unless $R^\ast \ll R_{\rm c}$. This does
not exclude the possibility of absorption by pair production with 
optical or UV photons. It should be noted that the conclusions we have derived 
are not based on the assumption of a specific radiation process for the 
TeV scale \grsn.

\begin{figure} [tb]
{\includegraphics[width=8.5cm]{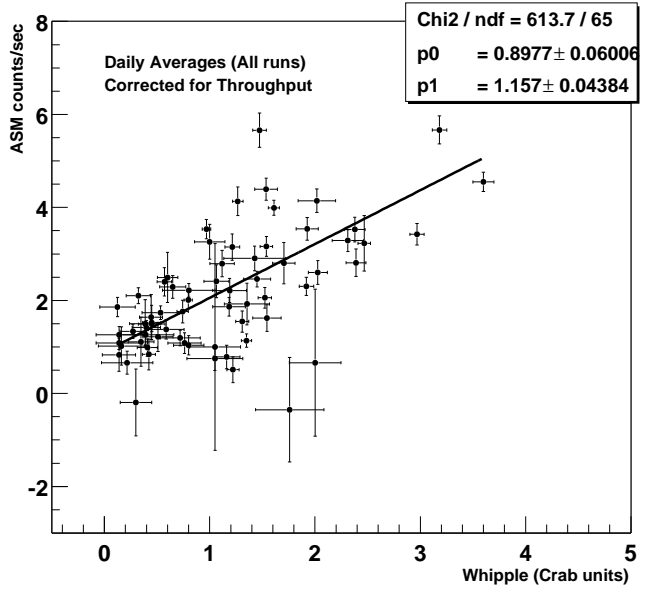}}
\caption{The correlation of TeV \gr rates from Mkn 421 with one-day
ASM quicklook count rates in the energy range of 2--12 keV
(taken from \citet{hold-ic}). The data shown have been taken between November 
2000 and April 2001.}
\label{holder-fig}
\end{figure}

Let us now return to the question whether or not a single, correlated 
outburst can 
be taken as evidence for a correlation between different wavebands. 
Fig.\ref{fegan-fig} shows the X-ray and \gr light curves of Mkn 421 for
a week in the year 2000. The X-ray light curve is essentially continuous,
except for two detector dropouts, one of which occurred at the time of 
the peak in \grsn. The \gr measurements have taken place only during the night,
for an imaging atmospheric \v Cerenkov telescope has been used. 

The variability in X-rays appears to be fairly well resolved. In a Fourier
spectrum of the light curve most of the power would reside at time scales of
10--20 hours and very little at smaller time scales. In \grs that is obviously 
different: most of the power in a Fourier spectrum would reside at time scales
around one hour and very little at longer time scales, except perhaps for
the outburst on day 5. Clearly, there is no one-to-one correlation between
X-rays and TeV \grsn. 

\citet{hold-ic} have compared the WHIPPLE light curve for Mkn 421 between
November 2000 and April 2001 with the RXTE ASM light curve of keV scale X-rays.
Fig.\ref{holder-fig} shows a scatter plot of the respective counts rates. 
Obviously, a linear regression provides a pretty bad fit. There is certainly a trend, that on days with a high \gr rate the X-ray rate is also enhanced,
but the relative scaling varies quite a lot. I don't know what the underlying
process is.
But I do know what the underlying process is not: it is not 
synchrotron-self-Compton scattering in a homogeneous source.
This implies that all deductions of physical parameters based on the 
assumption of a simple SSC model are of limited value, for the model doesn`t
apply.

\subsection{The TeV scale \gr spectra of AGN}

\begin{figure} [tb]
{\includegraphics[width=8.5cm]{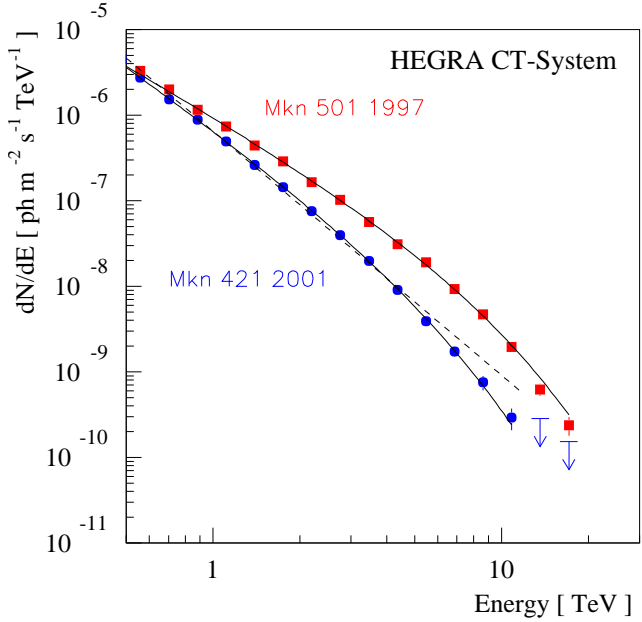}}
\caption{The Jan/Feb 2001 time-averaged Mkn 421 \gr spectrum as observed with HEGRA as well as the 1997 time-averaged spectrum of Mkn 501 (taken from 
\citet{kohnle-ic}). The solid line shows the fit of a power law with 
exponential cut-off, the dashed line shows a single power law fit. Upper limits
are 2$\sigma$ confidence level.}
\label{kohnle-fig}
\end{figure}

Earlier measurements indicated that the TeV \gr spectrum of Mkn 501 is 
curved, possibly caused by absorption, and that the spectrum of Mkn 421 up to
\gr energies around 10 TeV is
well described by a single power law. That was disturbing, because at 5 TeV or higher an effect of absorption on the infrared background should have been 
visible on account of lower limits on the infrared photon density in 
intergalactic space. Of all imaging atmospheric \v Cerenkov telescope 
operational to date, HEGRA offers the best energy resolution. Recent measurements have shown, that also the Mkn 421 \gr spectrum displays curvature
\citep{kohnle-ic}. Fig.\ref{kohnle-fig} clearly shows that the spectrum observed
in early 2001 cannot be represented by a single power law. The cut-off energy 
for a power law with exponential cut-off is $E_0 =(4.2-0.4+0.6)\ $TeV, the power law index is $s=2.33\pm 0.08$. This curved spectrum is compatible both with
data taken earlier and the results of WHIPPLE and CAT. The best fit to the
\gr spectrum of Mkn 501 is obtained for $E_0 =(6.2\pm 0.4)\ $TeV and $s=1.92\pm 0.03$. The power law index and the cut-off energy are statistically not 
completely independent of each other. Nevertheless, it is clear that the 
TeV scale \gr spectrum of Mkn 501 is harder than that of Mkn 421. 

\begin{figure} [tb]
{\includegraphics[width=8.5cm]{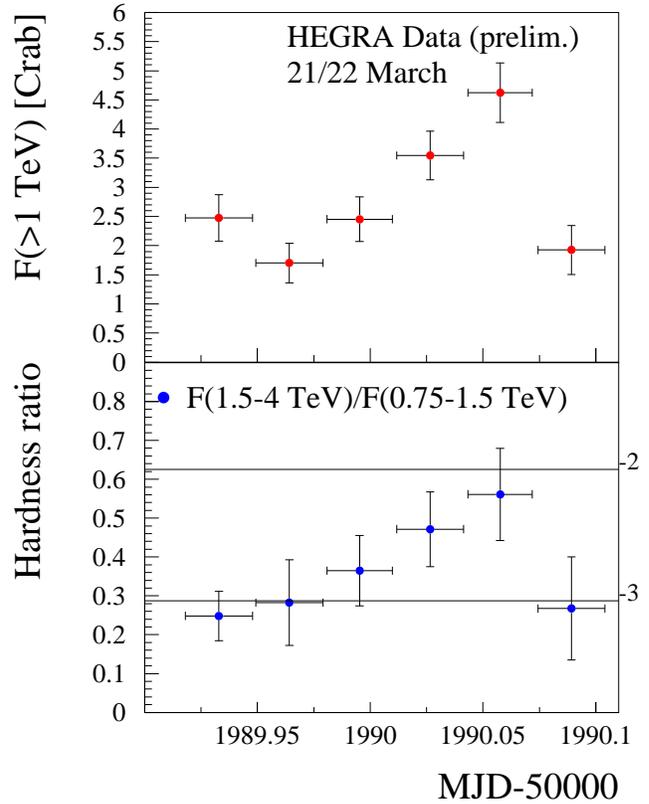}}
\caption{The TeV \gr rate from Mkn 421 and the hardness ratio observed with the
HEGRA array during the night of March 21/22, 2001
(from a viewgraph presented by D. Horns). 
The \gr spectrum is apparently harder when the count rate is high.}
\label{horns-fig}
\end{figure}

For models of \gr production in AGN it usually is an easy exercise to reproduce
the spectrum observed at a given time. Also, the observed time scales of
the variability are not problematic for the theories. However,
the models make fairly different predictions for how flares would evolve
at different \gr energies.  Clearly, what has been lacking so
far as constraint for the theories are observations of spectral variability.
This may change in the near future. Preliminary though the analysis is,
at this conference the HEGRA team has 
presented the first credible evidence for a hardening of the \gr spectrum
of Mkn 421 during an outburst (see Fig.\ref{horns-fig}).

\subsection{\gr absorption by the infrared background}

High energy \grs can interact with ambient radiation and form 
an electron/positron pair 
\begin{equation}
\gamma + \gamma \rightarrow e^+ + e^-
\label{9}
\end{equation}
The electrons would also be highly relativistic and would emit \grs at
energies somewhat smaller than the energy of the primary \gr that has produced
the pair. The secondary \grs would be emitted at a small angle with respect
to the primary $\gamma$-ray, even if the electron was not 
significantly deflected by
magnetic fields. Essentially, the $\gamma$-radiation cascades to lower energies
and at the same time is scattered out of the line-of-sight. For the
\gr flux from a point source this process corresponds to an absorption, with 
the radiation energy reappearing in the form of diffuse emission.

\begin{figure} [t]
{\includegraphics[width=8.5cm]{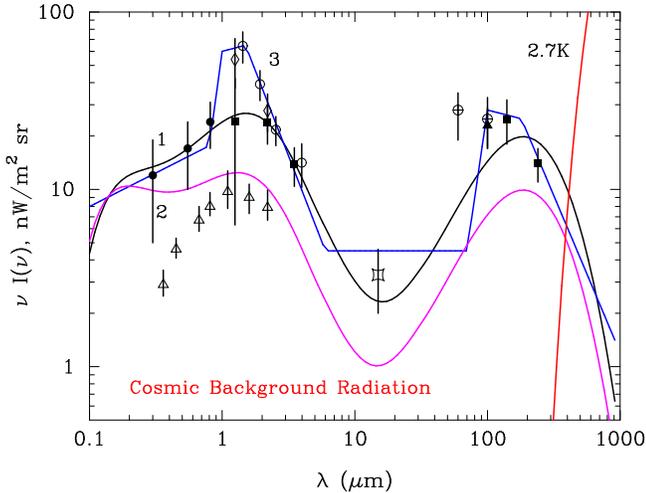}}
\caption{Three test spectra of the infrared background radiation shown in comparison with actual measurements (from a viewgraph presented by N. 
G\"otting). The triangles are effectively lower limits
from resolved sources, as is the point at 15 $\mu$m. The other data are from 
absolute photometry and may contain foreground emission which has not been
properly subtracted.}
\label{debra-fig}
\end{figure}

\begin{figure} [tb]
{\includegraphics[width=8.5cm]{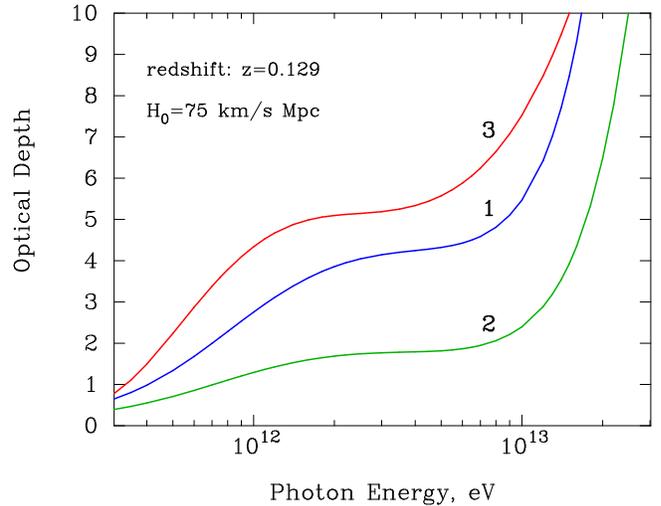}}
\caption{The optical depth for \grs as a function of energy for the three
model spectra of the infrared background shown in Fig.\ref{debra-fig} 
(from a viewgraph presented by N. G\"otting). The assumed redshift of the 
source is that of 1426+428.}
\label{14tau-fig}
\end{figure}

The pair production rate for an isotropic distribution of soft target photons
peaks at a few times the threshold energy, defined by ${E_\gamma}_1 {E_\gamma}_2
=0.25 m_e^2 c^4$, and falls of rapidly for higher interaction energies.
The target photons responsible for the absorption of TeV scale radiation are
thus in the infrared range. Fig.\ref{debra-fig} shows models and data
of the extragalactic infrared background light (for a comprehensive review 
see \citet{prim00}). Apparently, our knowledge of the actual intensity of the infrared background is accurate only to a factor of around two, depending on wavelength.
The HEGRA team has used three possible test spectra of the infrared background 
to estimate the optical depth for TeV \grs from AGN and the uncertainty thereof.
Spectrum 1 follows closely the model of \citet{prim00} for a 
Kennicut-IMF (initial mass function of stars). Spectrum 2 is close to the
lowest intensity allowed by the actual data. Spectrum 3 has been devised to reproduce recent measurements of very intense near-infrared background emission
\citep{matsu00,camb01}. Fig.\ref{14tau-fig} shows the optical depth thus determined for the recently detected BL Lac 1426+428 for the three model spectra
of the infrared background light.

Rather than correcting the observed \gr spectra of AGN for the effect of 
absorption, one can use models of the intrinsic \gr spectra, calculate the \gr
spectra after attenuation by the cosmic background radiation, and compare those
with the measured spectrum. Two groups (G\"otting for the HEGRA team, and
Vassiliev
for the VERITAS collaboration) have independently presented such calculations,
one assuming an intrinsic spectrum following a power law of arbitrary index, 
the other one assuming a power law with a fixed index of 1.92 based on an SSC
origin of the \grsn. 

\begin{figure} [tb]
{\includegraphics[width=8.5cm]{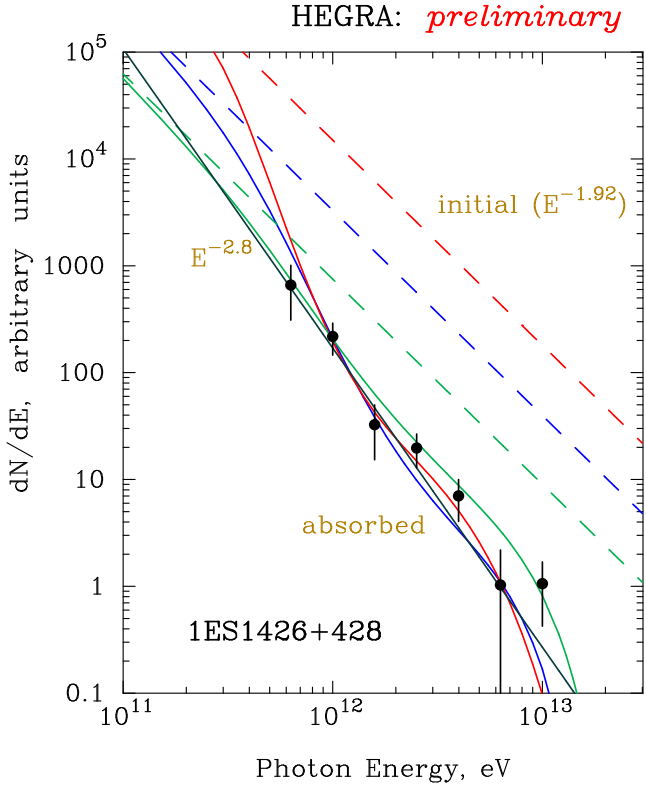}}
\caption{The preliminary spectrum of 1426+428 as measured with the HEGRA array
in comparison with model spectra (from a viewgraph presented by N. G\"otting). 
For the three models of the infrared background radiation shown in 
Fig.\ref{debra-fig} the attenuation of an assumed intrinsic power law \gr spectrum is calculated. Apparently none of the three models is in conflict with
the data.}
\label{14tev-fig}
\end{figure}

At this conference the HEGRA team has presented a preliminary \gr spectrum of
1426+428, which appears to be surprisingly well defined given the statistical 
significance of the detection as such. As shown in Fig.\ref{14tev-fig},
none of the three models of the cosmic infrared background is in conflict with
the data. Preliminary though they are, the results suggest that the 
intrinsic \gr spectrum
of 1426+428 is harder than a power law $E^{-2}$ up to about 10 TeV. Clearly,
a better defined \gr spectrum must be measured before definitive conclusion 
can be made. Nevertheless, with the confirmed detection of an AGN at a 
redshift of $z\simeq 0.13$ meaningful studies of the
cosmic infrared background radiation become feasible by means of TeV \gr 
astronomy.

\begin{figure} [tb]
{\includegraphics[width=8.5cm]{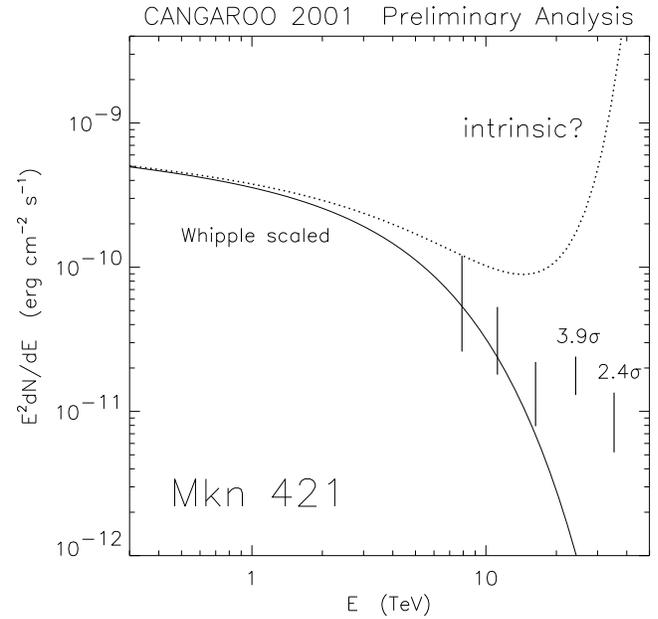}}
\caption{The Mkn 421 \gr spectrum observed at large zenith angle with CANGAROO II after preliminary analysis (according to a viewgraph presented by K. 
Okumura). The scaled average spectrum observed with WHIPPLE in 2001 is shown as the solid line for comparison. 
Given the high optical depth for photon energies beyond 20 TeV,
the intrinsic spectrum would have to look roughly like the dotted curve, if the
two data points at the highest energy were correct.}
\label{okumura-fig}
\end{figure}

Fig.\ref{14tau-fig} indicates that the optical depth for \grs increases rapidly
for photon energy beyond 10 TeV, whatever the actual model of the infrared
background. Even for Mkn 421, which resides at a quarter of the distance
of 1426+428, the optical depth at 20 TeV would be $\tau_2\simeq 1.3$ for spectrum 2 and $\tau_{1,3}\approx 5$ for the spectra 1 and 3 of the cosmic background radiation. Measurements at energies around 10 TeV or higher
can be performed with the imaging atmospheric \v Cerenkov telescopes, when
the source is far from the zenith, for both the threshold energy and the
effective area are then much higher than near zenith. Such measurements 
have recently
been performed with CANGAROO II, and the result of a preliminary analysis is 
shown in Fig.\ref{okumura-fig}. Up to the highest data point, which represents
\grs around 35 TeV, no effect of attenuation is visible in the spectrum, though
the optical depth should be much higher than unity for all possible models of 
the infrared background radiation. Consequently, the intrinsic \gr spectrum
of the source would have to be extremely hard beyond 20 TeV, if the
two data points at the highest energy were correct. The analysis of the
CANGAROO data is still preliminary, and thus the results may change beyond what
is indicated by the error bars. These results can be best summarized by stating
that they represent either a problem with the data analysis or a scientific 
sensation, for something would have to be seriously wrong in our understanding
of the universe, may it be Lorentz-invariance or the relation between redshift 
and distance or something else.
 
\section{Gamma-ray bursts}

Several models of Gamma-ray bursts (GRBs) predict TeV scale radiation
from inverse Compton scattering or other processes with comparable fluence
to the well measured MeV scale radiation (e.g. Dermer, this volume).
Measuring the VHE component of GRBs may be critical to the understanding
of the charged particle acceleration. However, the detection of TeV emission
from GRBs is complicated by the attenuation of VHE photons by interaction
with the intergalactic infrared radiation, for which the optical depth
is around unity for a redshift of $z=0.1$ at TeV \gr energies.

Sensitive though the atmospheric \v Cerenkov telescopes are, their 
field-of-view and duty cycle are too small to provide a good coverage of the
prompt emission from GRBs
detected by other resources such as BATSE. The air shower arrays
are much better suited to search for TeV emission from known GRBs. However,
the INCA and TIBET arrays have not found a \gr signal coincident with
BATSE bursts \citep{amenom-ic,cast-ic}. It should be noted that the altitude of 
INCA and the use of the "single particle" technique have allowed to work 
with a detection threshold of a few GeV. MILAGRO has conducted a search for
GRBs without any prior knowledge of the bursts position in the sky, its start
time and duration, which yielded no detection \citep{smith-ic}.
Given the temporal and spectral coverage of the searches performed so far, 
the one \gr excess coincident with a BATSE burst that was found by MILAGRITO
\citep{atkins00},
the smaller prototype of MILAGRO, has a significant probability of having
occurred by chance.

It is usually presumed that the afterglow emission of GRBs is caused by the sweep-up of interstellar matter by the decelerating relativistic blast wave.
\citet{meli-ic} have numerically investigated shock acceleration of particles
in this environment. They find that for highly relativistic blast waves 
($\Gamma \ge 100$) structured particle spectra would be produced, which
significantly differ from power laws. If $\Gamma \sim 1000$ proton acceleration
to $\sim 10^{20}\ $eV could be possible, which, if the protons would escape
from the system without loosing their energy, could be one possible source of 
ultra-high energy cosmic rays \citep{viet95,wax95}.

\section{Summary}

To date, a wealth of new exciting data is available in \gr astronomy, for a number of observatories using the imaging atmospheric \v Cerenkov technique
are operational and provide a very good flux sensitivity per source. The
prospects for the future are equally bright: four new imaging atmospheric 
\v Cerenkov telescope are under construction, which will allow observations
with a better flux sensitivity and a lower energy threshold than possible to date. With the advent of the forthcoming satellite-based GLAST experiment in a 
few years from now coordinated measurements at \gr energies between 50 MeV
and 20 TeV will be possible.

The data available to date have considerably furthered our understanding of the
high energy sky.

\begin{itemize}
\item Measurements of diffuse TeV scale \gr emission start to constrain models,
in particular those devised to explain the GeV excess in diffuse galactic
\gr emission.
\item There is still no unambiguous evidence of cosmic ray nucleon acceleration
in SNR or other possible sources of galactic cosmic rays.
\item The accuracy of the measurements is such that studies of spectral evolution during short-time \gr outbursts of AGN become feasible, thus 
constraining models of particle energization in these objects.
\item TeV \gr emission has been observed from a number of AGN ranging from
$\sim$0.03 to $\sim$0.13 in redshift, thus allowing to commence
studies of the infrared 
background light by disentangling the intrinsic \gr spectra of AGN and their
modification by \gr absorption through pair production in intergalactic space.
\end{itemize}

\begin{acknowledgements}

Partial support by the Bundesministerium f\"ur Bildung und
Forschung through DLR, grant 50 QV 0002, is acknowledged. A many colleagues
have helped by making available their results in electronic form, in some cases 
when those were still preliminary. 
Some results presented at the conference 
were obtained after the submission deadline for contributed papers. It is possible that my recollection of results presented only as talks or posters
and not as papers is incomplete. Therefore I would like to apologize to everyone who finds his work misinterpreted in this rapporteur paper. I did my very best.

\end{acknowledgements}

\end{document}